\documentclass[galaxies,article,accept,oneauthor,dvipdfm,12pt,a4paper]{mdpi} 
\lastpage{43}
\doinum{10.3390/galaxies1010031}
\pubvolume{1}
\pubyear{2013}
\history{Received: 15 April 2013; in revised form: 22 May 2013 / Accepted: 23 May 2013 / \\
Published: 29 May 2013}
\graphicspath{{fig/}}
\usepackage{graphicx,soul}

\def\a{\alpha}

\def\l{\lambda}

\def\r{\varrho}

\def\bm#1{\mbox{\boldmath{$#1$}}}

\def\ra{\rightarrow}

\Title{ Halo Models of Large Scale Structure 
and Reliability of Cosmological ${\bm N}$-Body Simulations
}

\Author{ Jos\'e Gaite
}
\address[1]{%
  {Instituto Universitario de Microgravedad IDR, ETSI Aeron\'auticos,
    Universidad Polit\'ecnica de Madrid,} Pza.\ Cardenal Cisneros 3, Madrid
  E-28040, Spain; E-Mail:~jose.gaite@upm.es; Tel.:+34-91-336-6353; Fax:
  +34-91-336-6363 }

\corres{.}

\abstract{ Halo models of the large scale structure of the Universe are
  critically examined, focusing on the definition of halos as {\em smooth}
  distributions of cold dark matter. This definition is essentially based on
  the results of cosmological $N$-body simulations. By a careful analysis of
  the standard assumptions of halo models and $N$-body simulations and by
  taking into account previous studies of self-similarity of the cosmic web
  structure, we conclude that $N$-body cosmological simulations are not fully
  reliable in the range of scales where halos appear. Therefore, to have a
  consistent definition of halos is necessary either to define them as
  entities of arbitrary size with a grainy rather than smooth structure or to
  define their size in terms of small-scale baryonic physics. }


\begin{document}

\section{Introduction}

Halo models of the large scale structure of matter are now very popular, as
simple searches on the Internet show: for example, a Google search with the
three words ``halo model cosmology'' produces 4,680,000 results, and an ArXiv
search for ``halo model'' yields ``too many hits'' and recommends a more
specific search. Naturally, the halos to which halo models refer are dark
matter halos, initially introduced to model the invisible matter surrounding
galaxies. However, present halo models are concerned with the large scale
distribution of halos in space as well as with the distribution of matter
within a single halo. In this respect, the modern report on halo models by
Cooray and Sheth \cite{CooSh} traces the appearance of these models to 1952,
in a paper about the spatial distribution of galaxies written by
Neyman~and~Scott~\cite{NeySc}, where they argue that it is ``useful to think
of the galaxy distribution as being made up of distinct clusters with a {range
  of sizes}.'' Thus, Neyman and Scott propose that the statistical theory of
the galaxy distribution is simplified by separating the full distribution into
one part corresponding to the distribution of galaxies within clusters and
another corresponding to the distribution of cluster centers in space. In
particular, they favor ``quasi-uniform'' distributions of clusters and mention
the Poisson distribution. Of course, this hypothesis is not in accord with
modern ideas, in which the strong clustering of clusters plays a fundamental
role in the large scale structure of matter. When galaxy clusters are replaced
with dark matter halos and we consider the distribution of halo centers in
space, we have the basic halo model. This distribution is indeed assumed to be
non-uniform and the study of halo correlation functions is an important part
of halo models \cite{CooSh}.

At any rate, halo models are not sufficiently supported by observations of the
large scale structure of matter, inasmuch as
dark matter 
has not been observed directly and the indirect observations of it, through
gravity, are strongly model dependent. Actually, our knowledge of the dark
matter distribution is mainly due to the results of cosmological $N$-body
simulations. As collisionless cold dark matter is assumed to be the main
component of the cosmic fluid and its dynamics is very simple to simulate,
many $N$-body simulations with large $N$ have been carried out and the type of
structure to which they give rise is well studied. Halo models seem to adapt
well to this type of structure, since the particles (or bodies) tend to form
smooth distributions on small scales that one can associate with halos, and
these halos are, on larger scales, clustered in irregular distributions with
definite features, such as filaments. Therefore, there appear distinct halos
with a range of sizes, which make up the large scale structure of matter.

As the cold dark matter (CDM) dynamics is purely gravitational and does not
introduce any scale, one may ask what determines the range of sizes of
halos. This is one of the points we intend to unveil. In fact, the absence of
scales immediately suggests that the CDM distribution should be scale
invariant, namely, a fractal distribution. In fact, fractal models and halo
models of the large scale structure can be merged in a model of fractal
distributions of halos \cite{EPL}. However, the resulting model is actually a
{\em multifractal} model in which halos are characterized by point-like
singularities and, if the full matter distribution is statistically
homogeneous, halos consist of grainy rather than smooth mass distributions, of
arbitrary size. Point-like singularities can also be present at the centers of
smooth halos, but halos of this type have, in addition, smooth components and
definite sizes, in contrast with multifractal halos, in which both definite
sizes and smoothness are precluded by scale invariance and statistical
homogeneity.

Of course, multifractal singularities only appear in continuous matter
distributions, and $N$-body simulations amount to a discretization of matter
that breaks the scale invariance of CDM dynamics. The discreteness limitations
of cosmological $N$-body simulations have been studied by \mbox{Splinter {\em
    et al}.\ \cite{Splinter}}. Their conclusions are very relevant to our
problem and are reproduced at the end, after examining the definition of halos
in cosmological $N$-body simulations.

In summary, our main concern is to find out if halo models of the large scale
structure are well justified, specifically, if smooth halos with a range of
sizes are well justified by cosmological $N$-body simulations. Regarding this
problem, we have to assess the spherical collapse and virialization model that
is supposed to lead to the formation of halos. The large scale distribution of
matter, whether made up by halos or not, displays definite features, namely,
filaments and walls, which constitute the famous ``cosmic web''
structure. This structure is reproduced by the adhesion model, which is worth
studying with regard to the formation of halos. Taking account of the
conclusions from our study of the spherical collapse and adhesion models, the
results of $N$-body simulations are reconsidered, to establish the role of the
breaking of scale invariance in them and its consequences for halo models. At
the end, some critical conclusions are presented and discussed.

\section{The Spherical Collapse Model and Virialization}

The spherical collapse model is a simplified model of gravitational collapse
that is supposed to give rise to the simplest type of halo germs. Its main
interest is that the spherical collapse, namely, the collapse of an initial
matter distribution that is spherically symmetric, is soluble, in the sense
that it consists of the one-dimensional gravitational dynamics of spherical
shells. In particular, let us consider, in a spatially flat
Friedmann--Lemaitre--Robertson--Walker universe, the collapse of an initial
``top-hat'' overdensity, namely, a sphere with constant density slightly
larger than the background density. Its evolution has a straightforward
solution in Lagrangian coordinates, and it undergoes several stages:
\linebreak(i) an initial expansion that follows the Hubble expansion at a
slower rate; (ii) as the expansion decelerates, the ``top-hat'' overdensity
reaches a maximum size and begins to contract; (iii) then, it collapses and,
if it stays spherically symmetric, its size tends to zero, but in practice it
is supposed to virialize and stabilize at some non-zero size.

Thus, the spherical collapse model leads to the formation of what one may call
a spherical halo, but the model {\em per se} does not prescribe its size. It
just assumes that a different process, namely, ``virialization'', takes over
at the end and produces an object of a definite size. On the other hand, there
is no way to predict this size, so the virialized object is {\em supposed} to
be a spherical object with a radius that is precisely {\em one half} of the
turn-around radius. This has the advantage of linking the sizes of the final
virialized halos to the initial spectrum of linear overdensities. On the other
hand, this link may look suspicious, because virialization embodies the
nonlinear and {\em chaotic} nature of gravitational dynamics, and chaos
implies erasure of initial conditions. Therefore, it is necessary to look into
the meaning of virialization in some~detail.

\subsection{Virialization}

Naturally, the stage of contraction in the spherical collapse of a uniform
sphere produces homologous spheres of decreasing size and increasing density
until reaching zero size and infinite density. This is as true for CDM as for
a gas, assuming that the process is adiabatic, namely, that the entropy does
not increase in it. A point of infinite density is a singularity, but one can
predict, under the assumption of reversibility, that it is followed by a
rebound and an expansion, until the sphere's radius gets back to the
turn-around radius. Therefore, the motion is oscillatory. However, this
reversibility is not realistic and one must expect irreversibility and entropy
growth, and, plausibly, the formation of a stable state of smaller size. This
stable, collapsed state must fulfill the (scalar) virial theorem, namely, {$E
  = -K + 3PV$}, where $E$ is its total energy, $K$ its (average) kinetic
energy, $V$ its (average) volume and $P$ the external pressure, which vanishes
in the ``top-hat'' collapse model. On the other hand, the reversible
oscillatory motion also fulfills the virial theorem, so ``virialization'' is
actually a misnomer. In fact, there is no way in which the virial theorem can
select a preferred size for the stable state. This stable, collapsed state is
rather the consequence of the type of processes known as ``violent
relaxation'' (redistribution in a rapidly varying gravitational potential) or
``chaotic mixing'' (exponential spreading of trajectories in phase space). We
must consider these processes to unveil how the collapse proceeds.

Of course, relaxation and mixing take place because the initial ``top-hat''
overdensity cannot be taken totally uniform and must contain density
perturbations inside. The inner overdensities (or underdensities) evolve and
grow just as the total overdensity does. Therefore, the spherical symmetry is
lost and the infalling particles do not converge to a point. Anyway, some
effects of non-uniformity can be studied within the spherical collapse model,
so that solubility in Lagrangian coordinates is maintained until
shell-crossing takes place. S\'anchez-Conde \emph{et al}.\ \cite{Sanchez-Conde}
undertake this study, after criticizing the standard assumptions of the
spherical collapse model, in particular the stabilization radius at one half
of the turn-around radius (``the justification \ldots is poor and lack a solid
theoretical background''). Their results do not support the common
assumptions, namely, the collapse factor $1/2$ and the time of
``virialization''. Presumably, the breaking of spherical symmetry makes the
common assumptions even less justifiable.

As a matter of principle, the characteristics of a stable state that has
undergone a relaxation process in which the thermodynamical entropy grows
cannot be determined by the initial conditions. Actually, entropy growth is
equivalent to loss of information, and the more entropy, the less information
about the process that has led to the stable state. Indeed, as we know from
thermodynamics, the most stable state is the one with the maximum entropy
allowed by the boundary conditions. In the gravitational case, the
maximum-entropy spherically-symmetric states are called isothermal
spheres. However, these are only local maxima of entropy, and there is no
global maximum. This is a consequence of the ``gravothermal catastrophe'': a
sufficient large central density tends to keep growing indefinitely (such an
isothermal sphere has negative specific heat). This shows, on the one hand,
that there can be temporary stable states of various sizes and, on the other
hand, that one must inevitably deal with singularities in the end.


At any rate, since the spherical collapse is unstable against non-radial
perturbations and, furthermore, cosmological $N$-body simulations show that
gravitational collapse is usually anisotropic and involves tidal interactions
with the surrounding matter, we consider next a more advanced model of
structure formation that includes these aspects.


\section{The Zeldovich Approximation and the Adhesion Model}

The Zeldovich approximation somewhat resembles the spherical collapse model,
insofar as it is soluble and indeed consists of a very simple dynamics in
Lagrangian coordinates, which holds until (non-spherical) shells
cross. However, the Zeldovich approximation, complemented by the adhesion
model, which gives a simple prescription for the dynamics after
shell-crossing, constitutes a more powerful and successful approach to the
formation of the large scale structure of the Universe
\cite{GSS}. Interestingly, the Zeldovich approximation implies that
``spherical collapse is specifically forbidden''~\cite{GSS}, because its
probability vanishes.

The Zeldovich approximation can be understood as the first order perturbative
approximation to the gravitational motion in Lagrangian coordinates
\cite{Lagrangian}; namely, the motion is given by ${\bm x} = {\bm x}_0 +
D(t)\,{\bm g}({\bm x}_0)$, where ${\bm x}$ is the comoving coordinate, ${\bm
  g}$ the peculiar gravitational field, and $D(t)$ the growth rate of linear
density fluctuations. Redefining time as $\tau=D(t)$, the motion is simply
uniform linear motion, with a constant velocity given by the initial peculiar
gravitational field. Naturally, nearby particles have different velocities,
and, as the linear solution is prolonged into the nonlinear regime,
trajectories cross at {\em caustic} surfaces, called ``Zeldovich pancakes'' in
this context \cite{GSS}. On the other hand, the formation of caustics is a
general feature of irrotational dust models, in Newtonian dynamics or in
general relativity, so it is reasonable to assume that caustics are indeed the
first cosmological structures.

\begin{figure}[H]
\centering{\includegraphics[width=9cm]{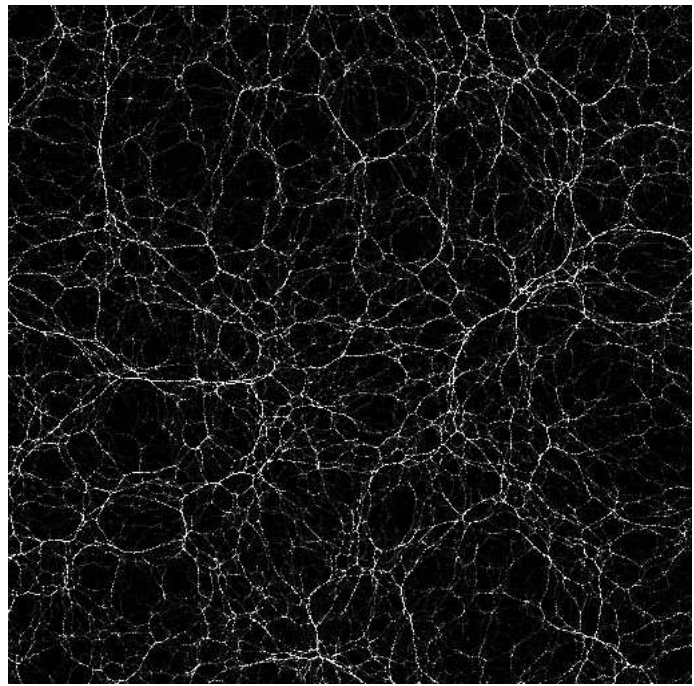}}
\caption{Cosmic web produced by the Burgers equation with random initial conditions.}
\label{cosmic-web}
\end{figure}

After a set of particles merge at a caustic, their subsequent evolution is
undefined. If no kinetic energy is dissipated ({\em adiabatic} collapse), the
particles cross (or rebound), like in a spherical collapse. Hence, if there
was no dissipation in caustics, there would be no real structure
formation. Therefore, the linear motion in the Zeldovich approximation is
supplemented with a {\em viscosity} term, resulting in the equation
\begin{equation}
\frac{d \widetilde{\bm u}}{d \tau} \equiv
\frac{\partial \widetilde{\bm u}}{\partial \tau} + 
\widetilde{\bm u}\cdot \nabla\widetilde{\bm u} =
\nu \nabla^2\widetilde{\bm u}
\label{Burg}
\end{equation}
where $\widetilde{\bm u}$ is the peculiar velocity in $\tau$-time. Let us
remark that dissipation and viscosity in CDM dynamics may not have the same
origin as in normal baryonic fluids \cite{GZ,BD}. To Equation~(\ref{Burg}),
it must be added the no-vorticity (potential flow) condition, $\nabla \times
\widetilde{\bm u} = 0$, implied by $\nabla \times {\bm g}({\bm x}_0) =
0$. Thus, Equation~(\ref{Burg}) is the three-dimensional form of the Burgers
equation for very compressible (pressureless) fluids \cite{GSS}. The limit
${\nu \ra 0}$ might seem to recover the caustic-crossing solutions but
actually is the high \linebreak Reynolds-number limit and gives rise to
Burgers turbulence. Whereas incompressible turbulence is associated with the
development of vorticity, Burgers turbulence is associated with the
development of {\em shock fronts}, namely, discontinuities of the
velocity. These discontinuities arise at caustics and give rise to matter
accumulation by inelastic collision of particles. The viscosity $\nu$ measures
the thickness of shock fronts, which become true singularities in the limit
${\nu \ra 0}$. This is the {\em adhesion model}, which produces, with the
appropriate random initial conditions, a characteristic network of sheets,
filaments and nodes, called ``the cosmic web'' \cite{GSS}. This distribution
of caustics is actually {\em self-similar}, with multifractal features
\cite{V-Frisch}. A simulation of the Burgers equation in the limit ${\nu \ra
  0}$ is shown in Figure~\ref{cosmic-web}.

One might think of identifying the nodes of the cosmic web with halos, but the
nodes produced by the adhesion model are just Dirac-delta singularities of
vanishing size. If ${\nu}$ is not zero, nodes have a size proportional to
${\nu}$. This size will be negligible if ${\nu}$ is identified with molecular
(baryon) viscosity, but it may be the right size if ``viscosity'' is due to
the mechanism proposed by Buchert and Dom{\'\i}nguez \cite{BD}. At any rate,
nodes are just one of the three types of singularities predicted by the
adhesion model, and the other types, namely, filaments and sheets, cannot be
identified with halos.

\section{$N$-Body Simulations}

$N$-body simulations of gravitational dynamics \cite{DR} have been very
helpful in the study of large scale structure formation and, in a way, have
been complementary to observations, since observations are biased towards the
baryonic matter, while $N$-body simulations take full account of the dark
matter, in particular, of non-baryonic matter. Collisionless non-baryonic CDM
is only subjected to gravitational forces, so it is fairly simple to simulate
its dynamics. Moreover, due to the advances in both hardware and software, now
it is possible to simulate the combined dynamics of CDM and baryon gas with
relatively good resolution. At any rate, the large scale dynamics is ruled by
the dominant component, namely, CDM. We employ the data from a large
simulation of CDM and gas carried out by the Mare Nostrum supercomputer in
Barcelona \cite{Gott1}. This simulation contains $1024^3$ dark matter
particles and the same number of gas particles in a comoving cube of 500
$h^{-1}$ Mpc edges. Later, we also employ, for a comparison, the CDM-only
Virgo Consortium GIF2 simulation, with $400^3$ particles in a 110 $h^{-1}$ Mpc
cube, as described by Gao \emph{et al}.\ \cite{GIF2}. Both simulations have
already been the object of multifractal analyses, by means of counts-in-cells
\cite{I1,I2}, and we can take advantage of the methods and results of these
analyses.

\begin{figure}[H]
\centering{\includegraphics[width=7.5cm]{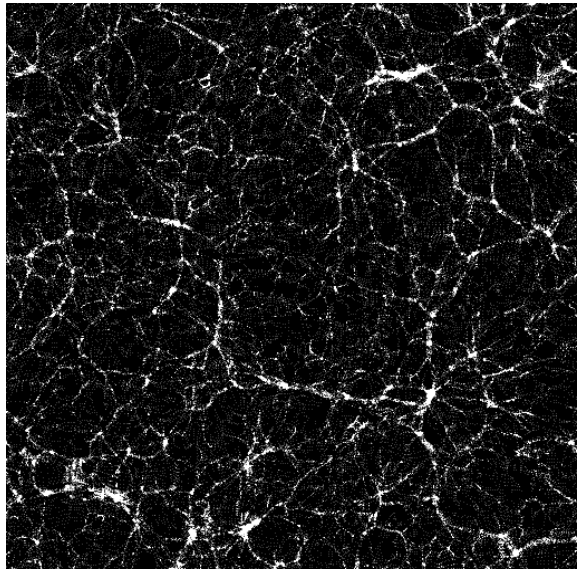}}
\caption{Dark matter slice of Mare Nostrum $N$-body simulation (cut off at $\r
= 4$).} 
\label{MN-slice}
\end{figure}

Naturally, we use the zero-redshift (present time) snapshots of either
simulation. A representative image of the matter distribution is given by the
distribution in a slice, see Figure~\ref{MN-slice}. This slice is prepared as
follows. First of all, we focus on the dominant CDM component of the Mare
Nostrum simulation. Since the number of particles is very large, it is useful
to coarse-grain the particle distribution to obtain a density representation
\cite{I1,I2}. The coarse-graining is carried out by using a mesh of cubes with
length such that the average density is one particle per cube \cite{I2}, so
the mesh-cube's edge is $1/1024$ of the simulation cube's edge. Furthermore,
given that the homogeneity scale is about 3\% of the simulation cube's edge
\cite{I2}, a quarter of a full slice is adequate to perceive the features of
the matter distribution (the lower left quarter is taken). In summary, our
slice consists of $512 \times 512$ mesh-cubes, and the density is given by the
number of particles in each one. To each mesh-cube corresponds a pixel in the
image, with an intensity proportional to the density in the mesh-cube. In
addition, in the slice represented in Figure~\ref{MN-slice}, the density field
has been cut off at $\r = 4$ ($\r = 1$ is the average density), so that the
contrast does not render invisible the pixels corresponding to low-density
cubes and one can appreciate the full cosmic-web structure. Indeed,
Figure~\ref{MN-slice} shows a self-similar structure that looks like the
structure in Figure~\ref{cosmic-web}.

\begin{figure}[H]
\centering{\includegraphics[width=7.5cm]{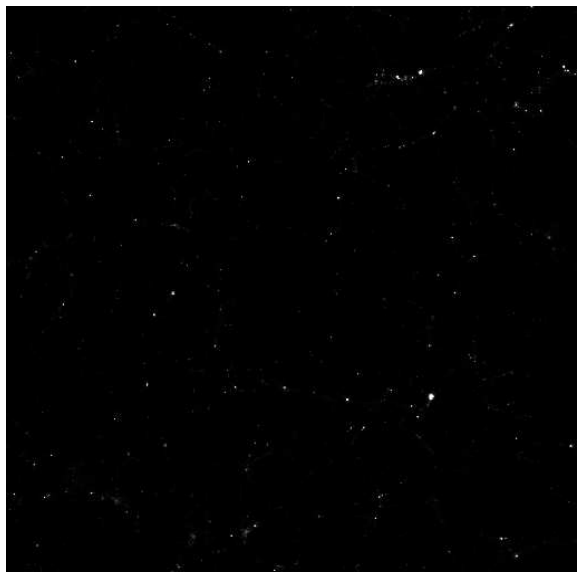}}
\caption{The same slice of Figure~\ref{MN-slice} but cut off at $\r = 256$. Notice the halos.}
\label{MN-slice_1}
\end{figure}

Nevertheless, one can wonder what is the appearance of the full density,
namely, including $\r > 4$. To see this, let us raise the cutoff to $\r =
256$, a value that is only exceeded by a few mesh-cubes and that, at the same
time, preserves some contrast in the picture. Amazingly, this change makes the
cosmic-web structure of Figure~\ref{MN-slice} vanish and the new image,
Figure~\ref{MN-slice_1}, resembles what one can see in a starry night, namely,
distinct bright spots with a (small) range of sizes. Naturally, these bright
spots must be identified with dark matter halos rather than with stars. To
understand the transformation from a cosmic web structure to a distribution of
halos of similar size, we must spell out the various scales that play a role
in cosmological $N$-body simulations.
Of course, the first scale is the simulation cube's edge, but we can refer the
remaining scales to it and, hence, assign it the value of unity. The next
scale is the discretization length $N^{-1/3}$, namely, the length of the edge
of the mesh cube such that there is one particle per cube on
average. Naturally, the mesh of these cubes is used for the counts-in-cells
and $N^{-1/3}=1/1024$ is the coarse-graining length (in physical units, $0.5\;
h^{-1}$ Mpc, in the Mare Nostrum simulation). There is another scale: the
gravity cutoff or softening length, which is necessary to avoid numerical
problems when two particles get close and the force between them gets too
large. The softening length is of the order of some Kpc, in particular, it is
$15\;h^{-1}$ Kpc in the Mare Nostrum simulation. (In the GIF2 simulation, the
discretization length is {$0.25\; h^{-1}$ Mpc and the softening length is $=
  7\;h^{-1}$ Kpc, so their ratio is almost the same.) Besides, there are other
  scales in the initial conditions, but we are only concerned with the scales
  in the dynamics. In summary, while CDM dynamics is scale free, we see that
  $N$-body simulations of it introduce two scales. Therefore, the appearance
  of halos in Figure~\ref{MN-slice_1} must be due to the presence of these
  scales, which prevent the formation of a truly self-similar cosmic
  web. Indeed, scale invariance can certainly be measured for scales between
  the homogeneity scale and the discretization scale \cite{I1,I2}. In
  addition, the halos in Figure~\ref{MN-slice_1} have sizes of the order of
the discretization length, which is the larger of the two scales.

While Figure~\ref{MN-slice} or Figure~\ref{MN-slice_1} show the matter
distribution between the homogeneity scale and the discretization scale, they
do not show the distribution on smaller scales, that is, they do not show what
one might call the mass distribution inside halos. The most populated
mesh-cube is located at the position $(380/512,159/512)$ in the slice and
contains $2466$ particles (to be compared with one per mesh-cube, on
average). Nearby cubes are overpopulated as well, so we define the heaviest
halo as the one formed by all of them together. To be precise, we choose $4
\times 4$ adjacent cubes of the slice and we display the (projected) particle
positions in them in Figure~\ref{halo}. Other halos in the slice have a
similar aspect. Patently, the matter distribution on these small scales is
very different from the cosmic-web distribution between the homogeneity scale
and the discretization scale: now we perceive a nearly smooth distribution
(this also happens for the GIF2 simulation, naturally). The smoothness of the
distribution inside halos is presumably due to the gravity softening. However,
the scale that seems to mark the transition from an irregular cosmic-web
distribution to a smooth distribution is the discretization scale. The
transition over this scale has an even more definite and sharp effect on the
statistics of halo masses, as we show next.

\begin{figure}[H]
\centering{\includegraphics[width=7cm]{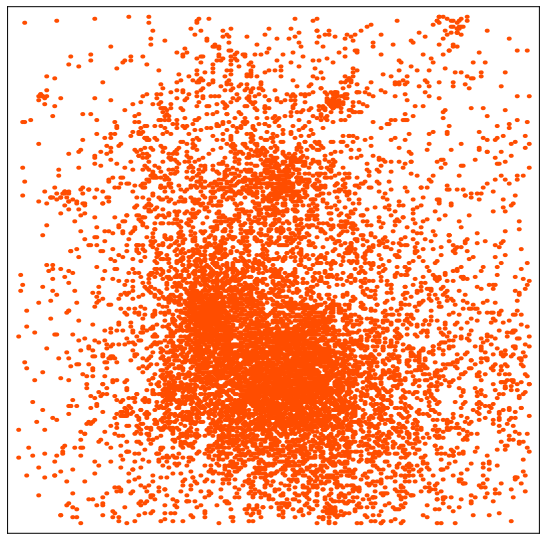}}
\caption{Zooming in on the largest halo: $4 \times 4$ pixels at position
  $(0.74,0.31)$ in Figure~\ref{MN-slice_1}.}
\label{halo}
\end{figure}

\subsection{Halo Mass Statistics}

Since we have seen that the halo sizes in $N$-body simulations are about the
discretization scale, it is appropriate to define halos so that they have
precisely this size, for the sake of simplification. Then, one can easily
measure by counts-in-cells the mass function of halos, namely, the number of
halos of a given mass. The halo mass functions of the Mare Nostrum or GIF2
zero-redshift snapshots follow very definite power laws precisely when the
halos have the size of the discretization scale \cite{I1,I2}. This is shown in
Figure~\ref{haloMF}, where the respective constant size halo mass functions
for variable size are displayed: power laws for size $N^{-1/3}$ are fulfilled
by all halos except the most massive ones. As the halo size increases, the
straight line bends, becoming convex from above, as expected in a multifractal
distribution \cite{I1,I2}. On the contrary, as the halo size decreases, the
straight line becomes concave from above, because the number of mesh-cubes
with few particles must then increase. Notice that the sizes chosen in the
GIF2 simulation are not exact multiples of $N^{-1/3}$, in consonance with the
characteristics of this simulation and our numerical methods: the GIF2
simulation contains $400^3$ dark matter particles and we use powers of 2
\cite{I1}.

\begin{figure}[H]
\centering{\includegraphics[width=4.5cm]{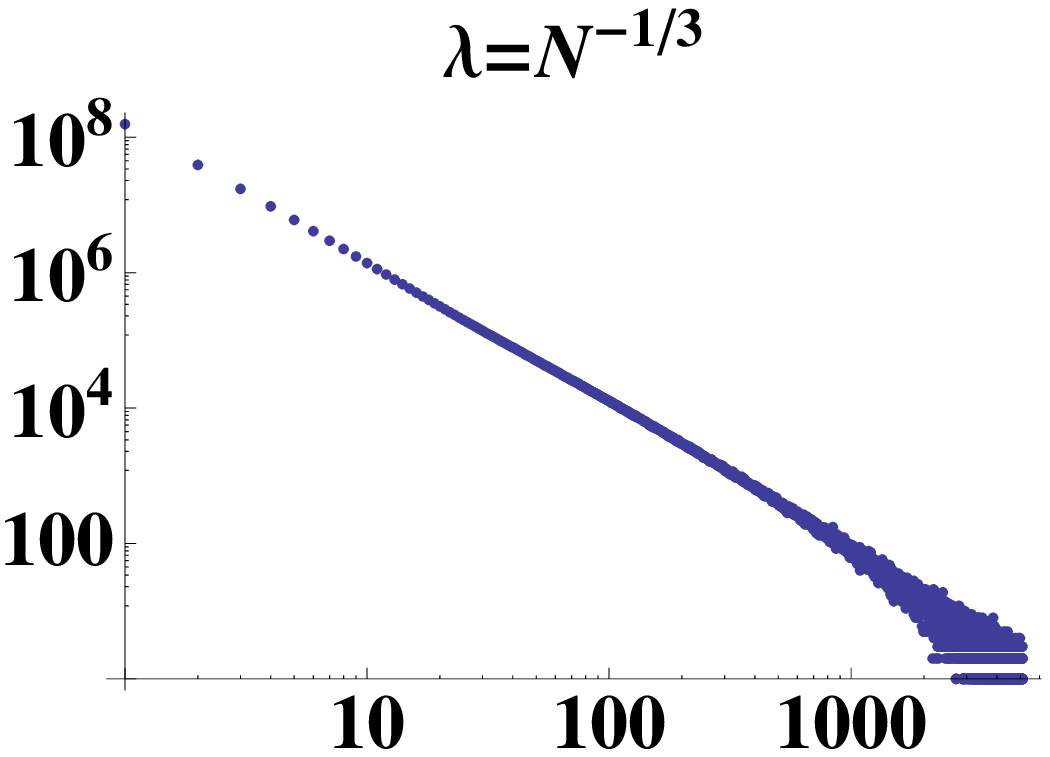}
\includegraphics[width=4.5cm]{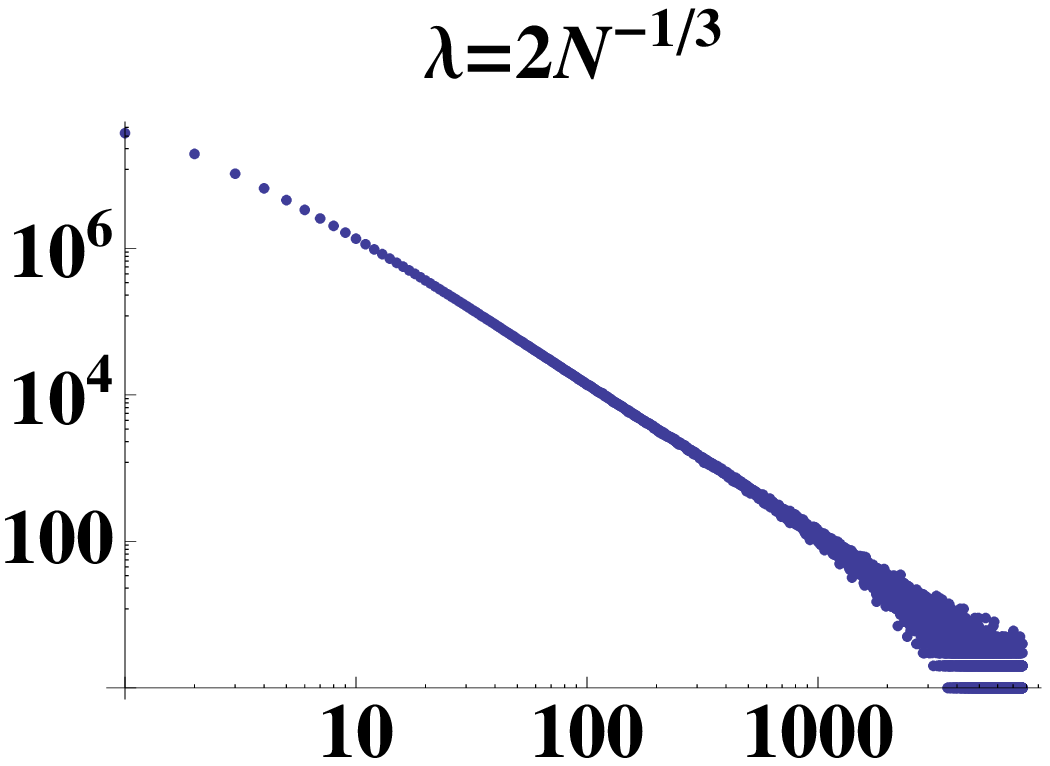} 
\includegraphics[width=4.5cm]{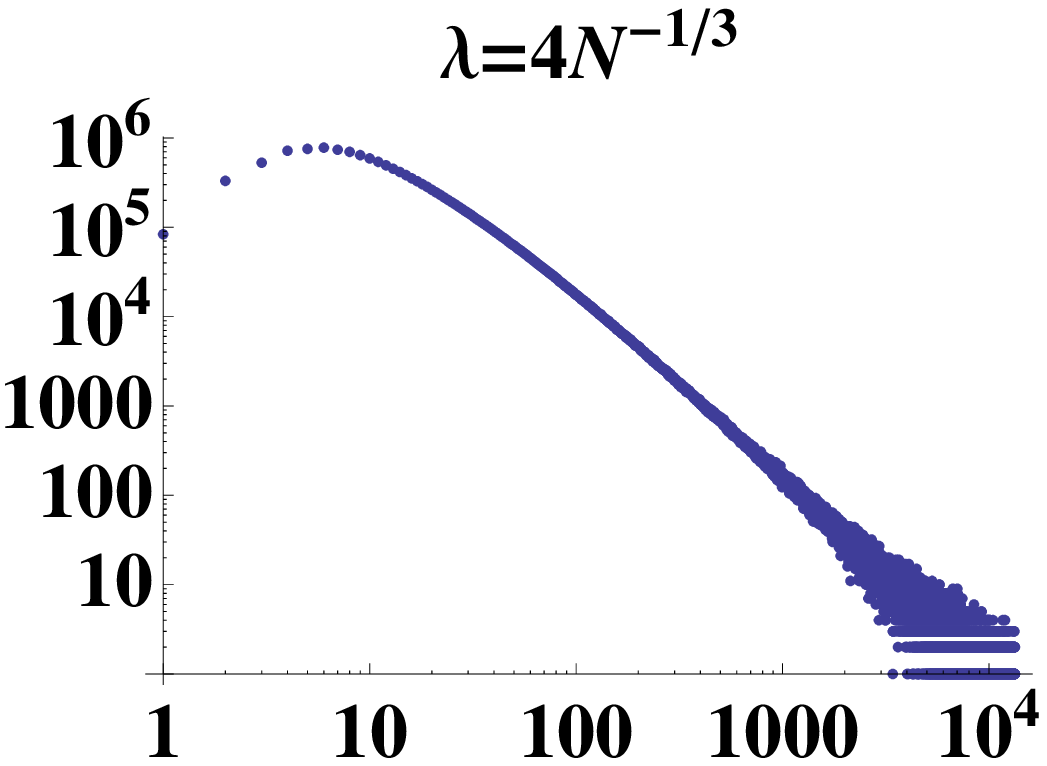} }
\\[4mm]
\centering{
\includegraphics[width=4.5cm]{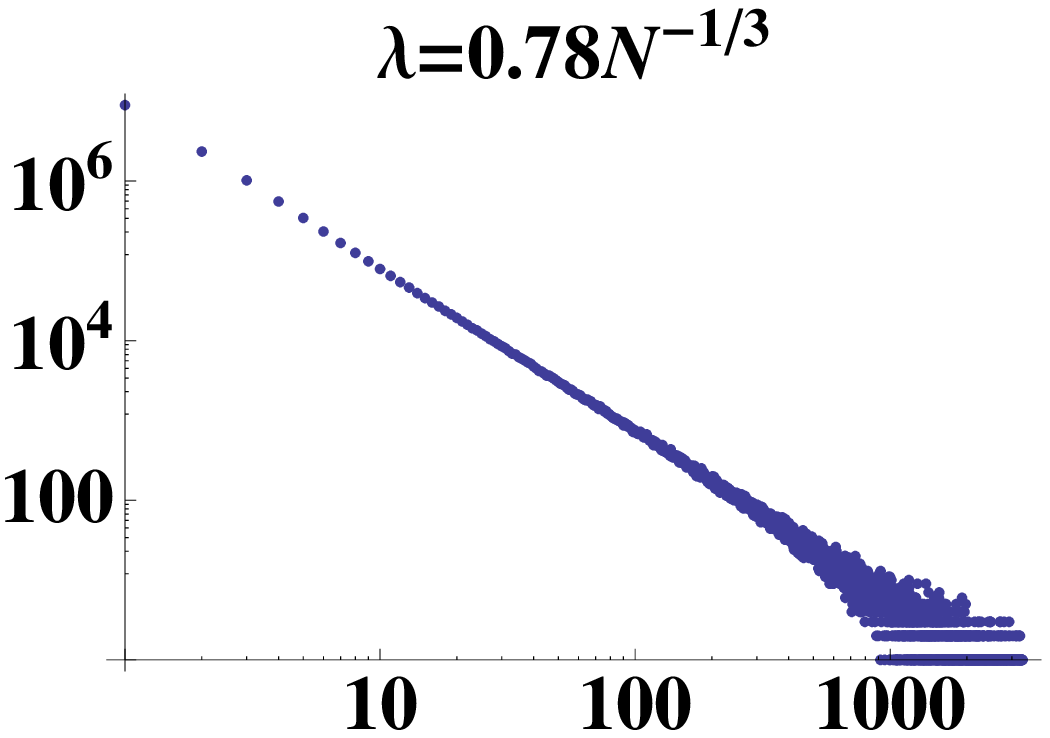} 
\includegraphics[width=4.5cm]{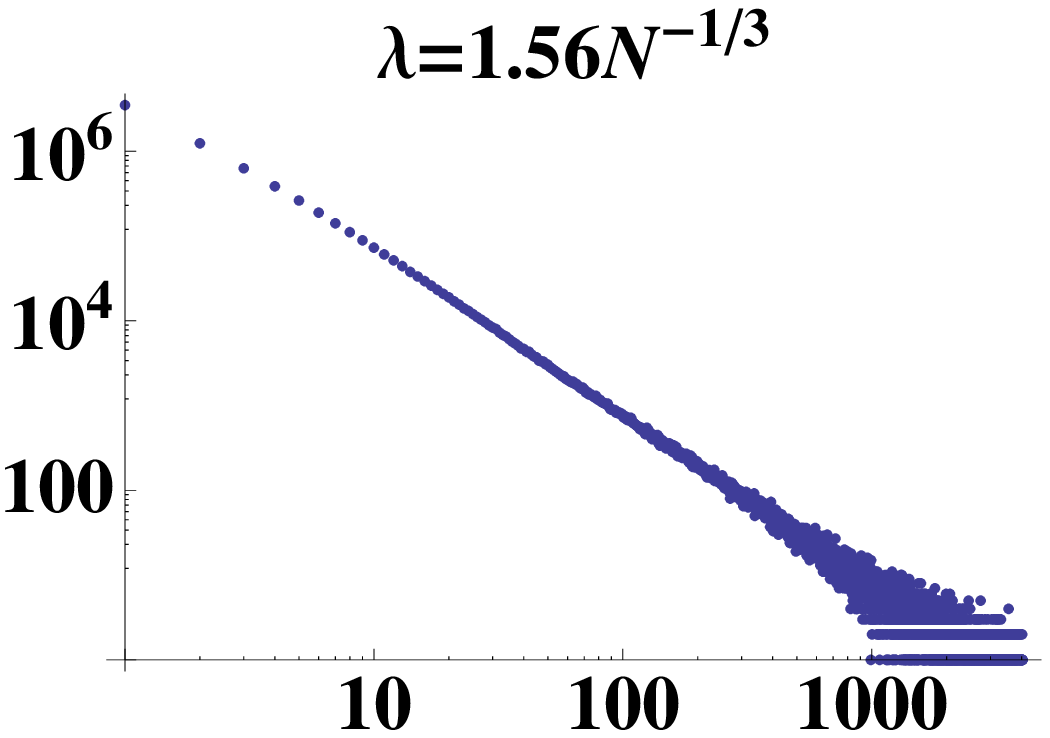} 
\includegraphics[width=4.5cm]{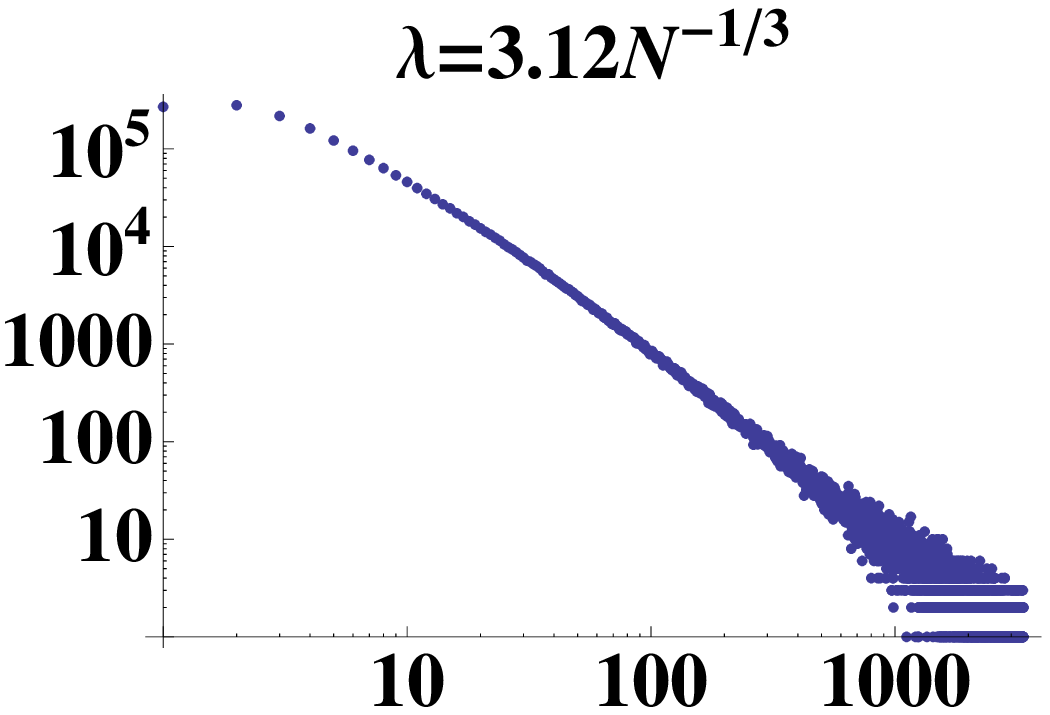} }
\caption{Constant size halo mass functions for variable size $\l$, for the
  Mare Nostrum, above, and GIF2, below, $z=0$ snapshots. Abscissas: halo mass;
  ordinates: number of~halos.}
\label{haloMF}
\end{figure}

The power-law mass function of halos at the discretization scale is found in
every cosmological \linebreak $N$-body simulation that we have analyzed,
besides the Mare Nostrum and GIF2 simulations, but we have no simple
explanation of it. It can be connected with the Press--Schechter theory of
structure formation by spherical collapse of overdensities in a Gaussian
distribution, but the power-law exponent is just beyond the allowed range
\cite{I1,I2}. In contrast, the parabola like shape (in a log-log plot) seen on
larger scales is explained by a lognormal like model that, in turn,
corresponds to a multifractal model of the matter distribution on those
scales. This model has been described in detail before (see \cite{I1,I2} and
references therein), so we now restrict ourselves to properties that are
relevant with regard to halos.  \newpage
\section{Scale Invariance and Halos}

Let us review very briefly the multifractal model of the large-scale structure
in cosmological $N$-body simulations, focusing on the CDM component of the
Mare Nostrum simulation. In the multifractal model, the coarse-grained density
$\r({\bm x},r)$ (defined by counts-in-cells or any suitable method), at the
point ${\bm x}$ and for coarse-graining length $r$, fulfills {$\r({\bm x},r)
  \sim r^{\a(}\mbox{\boldmath{$^x$}}^{)}{}^{-3}$}, where $\a \geq 0$ (see
\cite{Falcon} for a precise definition). Consequently, the point density
$\lim_{r\ra 0} \r({\bm x},r)$ is finite and non-vanishing only if $\a=3$,
while it is infinite for $\a < 3$, and zero for $\a > 3$. Therefore, it is
natural to associate points ${\bm x}$ such that $0 \leq \a({\bm x}) < 3$,
namely, density singularities, with halos and points such that $\a({\bm x}) >
3$ with cosmic voids. At any rate, multifractality is ensured by the power-law
behavior of the density with respect to the coarse-graining length or,
equivalently, by the power-law behavior of the statistical moments $M_q(r)$ of
the distribution \cite{Falcon}. A multifractal can be characterized by its
multifractal spectrum, namely, the fractal dimension $f(\a)$ of the set of
points ${\bm x}$ with exponent $\a$. Notice that the multifractal spectrum can
be defined for any distribution with singularities, not just for self-similar
distributions. However, multifractal spectra of self-similar distributions
have typical parabola like shapes \cite{Falcon,Halsey}.

One proof of multifractality consists in computing the coarse multifractal
spectrum for several coarse-graining lengths $r$ and showing that it does not
depend on $r$. We reproduce in Figure~\ref{MFspec} the coarse multifractal
spectrum of the dark matter in the Mare Nostrum simulation for {$r =
  \{1,2,4,8\}\times N^{-1/3}$}, which cover most of the scaling range
\cite{I2}. The extent of the scaling range and the transition to homogeneity
are better perceived in the scaling of moments $M_q(r)$ \cite{I2}. The scaling
range, which goes from the discretization length (or somewhat below) to the
homogeneity scale, extends over two decades, at the most.

\begin{figure}[H]
\centering{\includegraphics[width=7.5cm]{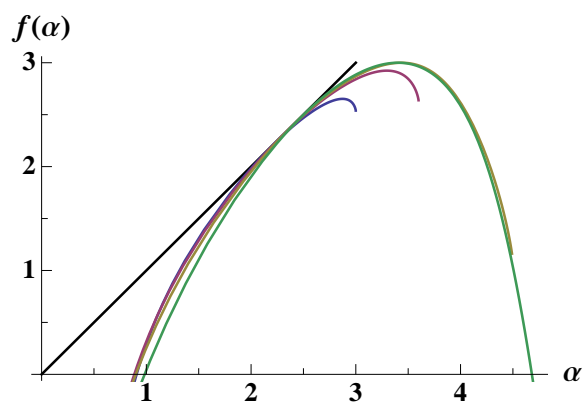}}
\caption{Multifractal spectra of the dark matter in the Mare Nostrum
  simulation, for \mbox{$r = \{1,2,4,8\}\times N^{-1/3}$} (blue, red, brown
  and green, respectively).}
\label{MFspec}
\end{figure}

Unfortunately, the scaling range in three-dimensional $N$-body simulations
cannot be very large, for the time being. In contrast, one-dimensional
cosmological $N$-body simulations with moderate $N$ can reach truly compelling
scaling ranges: the simulations of Miller {\em et al}.\ \cite{Miller}, with $N
\leq 2^{18} \simeq 2.6\times 10^{5}$, and of Joyce and Sicard \cite{Joyce},
with $N=10^5$, reach almost 4 decades! Moreover, the analysis by Joyce and
Sicard of the ``halos'' formed in their simulation has led them to state,
regarding three-dimensional halos, that CDM halos are ``not well modeled as
smooth objects'' and that ``the supposed `universality' of [halo] profiles is,
like apparent smoothness, an artifact of poor numerical resolution''. These
conclusions agree with the conclusions from our own analyses in three
dimensions \cite{EPL,I1,I2}: if we are to preserve the concept of CDM halos,
they are to be defined as grainy structures in a self-similar distribution
rather than smooth structures with a range of sizes.

A different but very interesting demonstration of, on the one hand, scale
invariance and of, on the other hand, the discreteness limitations of
cosmological $N$-body simulations is provided by the work of Gottl\"ober {\em
  et al}.\ \cite{Gott}. The purpose of their work is to assess the problem of
the emptiness of cosmic voids by means of $N$-body simulations. To do this,
Gottl\"ober {\em et al}.\ \cite{Gott} performed a low resolution simulation
and then resimulate voids with high resolution, namely, with the resolution
corresponding to replacing each particle with 512 particles. Naturally, they
find that the voids are no longer empty and, furthermore, the high-resolution
dark matter distribution inside large voids is such that ``haloes are arranged
in a pattern, which looks like a miniature universe.'' In other words,
Gottl\"ober {\em et al}.\ find that a higher resolution brings out in a void
the invisible structure below the discretization scale, demonstrating
self-similarity of the full structure. One can infer that a resimulation of
halos with higher resolution must bring out as well their grainy, self-similar
structure.


\section{Discussion and Conclusions}

It seems inevitable to conclude that the presence of an intrinsic scale in
cosmological $N$-body simulations, namely, the discreteness scale $N^{-1/3}$,
severely affects the type of mass distribution that is produced below that
scale, to the extent that the smooth halos with a range of sizes about that
scale that are commonly seen in these simulations are probably an artifact of
insufficient resolution.

The problems of cosmological $N$-body simulations below the discreteness scale
have already been noticed by Splinter {\em et al}.\ \cite{Splinter}: their
comparison of results of various $N$-body simulations reveals that ``codes
never agree well below the mean comoving interparticle separation'' (which is
another name for the discreteness scale). Therefore, one might think that the
smoothness of halos that is seen on these small scales should have been
questioned before. Probably, this has not occurred (or has had no
consequences) because of the popularity of the spherical collapse
model. However, now it appears that this model does not necessarily predict
{\em smooth} spherical halos and, in addition, its range of application is far
more restricted than usually assumed.

In fact, the adhesion model is more adequate than the spherical collapse model
to provide a general description of large scale structure, namely, to produce
the typical cosmic web structure perceived in both CDM simulations and
observations of the galaxy distribution. The cosmic web is self-similar, so
the adhesion model suggests that the size of halos or, in general, the size of
cosmic-web structures is determined by small-scale processes that can be
lumped into an effective viscosity that breaks the scale invariance. The
question is, of course, how such small-scale processes determine the scale at
which scale invariance is broken and how this scale compares with the
discreteness scale $N^{-1/3}~>~0.2\, h^{-1}$~Mpc~(generally).

First of all, let us remark that the real CDM is probably discrete. Indeed,
current models of CDM favor a WIMP composition. Neutralinos, for example, may
have a mass $< 1$ TeV. Therefore, comparing with the mass resolution of
cosmological $N$-body simulations (e.g., $8.24\times 10^9\,h^{-1}$~M$_\odot$
in the \linebreak Mare Nostrum simulation), there is such a huge factor
($10^{64}$ in the Mare Nostrum simulation) that the CDM distribution on
astrophysical scales is continuous, in practice, and, hence, $N$-body
simulations can in no way reproduce the real CDM dynamics on small scales.

We could also consider that the scale invariance of CDM dynamics is broken on
small scales by an aspect of it that is not taken into account by $N$-body
simulations: the collapse of CDM overdensities eventually produces densities
and velocities that make Newtonian physics invalid and require relativistic
physics. In general relativity, a mass has an associated length scale, namely,
its Schwarschild radius. Consequently, in the ``top-hat'' collapse model, for
example, there is an intrinsic scale, which, in contrast with the usually
assumed scale, is not arbitrary and, furthermore, is independent of the
initial conditions. Naturally, this new scale arises in connection with
supermassive black hole formation and, arguably, the size of these black holes
is not relevant on cosmological scales.

In conclusion, the CDM dynamics does not seem to generate any small scale that
is cosmologically relevant. Thus, it seems natural to either define a sort of
scale invariant halos \cite{EPL,I1,I2} or to turn to the baryon
physics. However, it is not easy to think of a definite scale in the baryonic
physics that marks the end of scale invariance. As a matter of fact, the gas
in the Mare Nostrum simulation follows the same scaling laws as the CDM does,
despite the presence of biasing \cite{I2}. At any rate, the modeling of
baryonic physics in cosmological $N$-body simulations is still in its infancy
\cite{DR}. What seems clear is that the standard conclusions about smooth
halos with a range of sizes drawn from state-of-the-art $N$-body simulations,
especially, CDM-only simulations, must be reassessed.



\bibliographystyle{mdpi}
\makeatletter
\renewcommand\@biblabel[1]{#1. }
\makeatother

\end{document}